\documentstyle[12pt,aaspp4]{article}
\begin{document}

\newcommand{\up}[1]{\ifmmode^{\rm #1}\else$^{\rm #1}$\fi}
\newcommand{\zdot}{\makebox[0pt][l]{.}}
\newcommand{\upd}{\up{d}}
\newcommand{\uph}{\up{h}}
\newcommand{\upm}{\up{m}}
\newcommand{\ups}{\up{s}}
\newcommand{\arcd}{\ifmmode^{\circ}\else$^{\circ}$\fi}
\newcommand{\arcm}{\ifmmode{'}\else$'$\fi}
\newcommand{\arcs}{\ifmmode{''}\else$''$\fi}

\title{The Araucaria Project. The Distance to the Local Group
Galaxy IC 1613 from 
Near-Infrared Photometry of Cepheid Variables.
\footnote{Based on  observations obtained with the NNT telescope 
at ESO/La Silla for programs 074.D-0318(B) and 074.D-0505(B)
}
}
\author{Grzegorz Pietrzy{\'n}ski}
\affil{Universidad de Concepci{\'o}n, Departamento de Fisica, Astronomy
Group,
Casilla 160-C,
Concepci{\'o}n, Chile}
\affil{Warsaw University Observatory, Al. Ujazdowskie 4, 00-478, Warsaw,
Poland}
\authoremail{pietrzyn@hubble.cfm.udec.cl}
\author{Wolfgang Gieren}
\affil{Universidad de Concepci{\'o}n, Departamento de Fisica, Astronomy Group,
Casilla 160-C, Concepci{\'o}n, Chile}
\authoremail{wgieren@astro-udec.cl}
\author{Igor Soszy{\'n}ski}
\affil{Universidad de Concepci{\'o}n, Departamento de Fisica, Astronomy Group, 
Casilla 160-C, Concepci{\'o}n, Chile}
\affil{Warsaw University Observatory, Al. Ujazdowskie 4, 00-478, Warsaw,
Poland}
\authoremail{soszynsk@astro-udec.cl}
\author{Fabio Bresolin}
\affil{Institute for Astronomy, University of Hawaii at Manoa, 2680 Woodlawn 
Drive, 
Honolulu HI 96822, USA}
\authoremail{bresolin@ifa.hawaii.edu}
\author{Rolf-Peter Kudritzki}
\affil{Institute for Astronomy, University of Hawaii at Manoa, 2680 Woodlawn 
Drive,
Honolulu HI 96822, USA}
\authoremail{kud@ifa.hawaii.edu}
\author{Massimo Dall'Ora}
\affil{INAF-Osservatorio Astronomico di Capodimonte, via Moiariello 16, I-80131 Naples, Italy}
\authoremail{dallora@na.astro.it}
\author{Jesper Storm}
\affil{Astrophysikalisches Institut Potsdam, An der Sternwarte 16, D-14482 Potsdam,
Germany}
\authoremail{jstorm@aip.de}
\author{Giuseppe Bono}
\affil{INAF-Osservatorio Astronomico di Roma, via Frascati 33, 00040 Monte Porzio Catone,
Italy}
\authoremail{bono@mporzio.astro.it}

\begin{abstract}
We have measured accurate near-infrared magnitudes in the J and K bands of 39 Cepheid variables
in the irregular Local Group galaxy IC 1613 with well-determined periods and optical VI light curves.
Using the template light curve approach of
Soszy{\'n}ski, Gieren and Pietrzy{\'n}ski, accurate mean magnitudes were obtained from
these data which allowed to determine the distance to IC 1613 relative to the LMC
from a multiwavelength period-luminosity solution in the optical VI and near-IR JK bands,
with an unprecedented accuracy. Our result for the IC 1613 distance is 
$(m-M)_{0} = 24.291 \pm 0.014$ (random error) mag, with an additional systematic uncertainty smaller
than 2\%. From our multiwavelength approach, we find for the total (average) reddening to
the IC 1613 Cepheids $E(B-V) = 0.090 \pm 0.007$ mag, which is significantly higher than the
foreground reddening of about 0.03 mag, showing the presence of appreciable dust extinction
inside the galaxy. Our data suggest that the extinction law in IC 1613 is very similar
to the galactic one. Our distance result agrees, within the uncertainties, with two earlier infrared Cepheid
studies in this galaxy of Macri et al. (from HST data on 4 Cepheids), and McAlary et al. (from ground-based
H-band photometry of 10 Cepheids), but our result has reduced the total uncertainty on
the distance to IC 1613 (relative to the LMC) to less than 3\%. With distances to nearby
galaxies from Cepheid infrared photometry at this level of accuracy, which are currently being
obtained in our Araucaria Project, it seems possible to significantly reduce the systematic
uncertainty of the Hubble constant as derived from the HST Key Project approach, by improving
the calibration of the metallicity effect on PL relation zero points, and by improving
the distance determination to the LMC.
\end{abstract}

\keywords{distance scale - galaxies: distances and redshifts - galaxies:
individual(IC 1613)  - stars: Cepheids - infrared photometry}

\section{Introduction}
Cepheid variables are the most important standard candles to calibrate
the first rungs of the extragalactic distance ladder, out to some 30 Mpc. As young
stars, Cepheids tend to lie in dusty regions in their spiral or irregular parent galaxies.
As a consequence, Cepheid distances derived from the period-luminosity (PL) relation
in optical photometric bands are quite sensitive to a precise knowledge of the total
reddening, foreground and intrinsic, of their parent galaxy. While the galactic 
foreground reddening towards any
direction in the sky is usually well established, particularly in directions far away
from the galactic equator, the correct assessment of the reddening produced by
dust extinction {\it intrinsic to the host galaxy} is usually a difficult task, and in most work
on Cepheid distances based on optical data such an intrinsic contribution to the
reddening has simply been ignored. Just for this one
particular reason, it is clear that more accurate Cepheid distances to galaxies 
can be derived in near-infrared passbands, where dust absorption is small as compared
to visual wavelengths, and the distance results become increasingly independent of errors
in the assumed total reddenings. Efforts along these lines have started in the early eighties 
with the pioneering work of McGonegal et al. (1982), and Welch et al. (1985).
Yet, an important obstacle to carry out accurate Cepheid distance work
in the infrared has been, until very recently, the lack of well-calibrated fiducial
PL relations in the near-infrared JHK bands. This problem has now been solved 
by the work of Persson et al. (2004) who provided such well-calibrated relations for the LMC. 
Very recently, Gieren et al. (2005a) have also provided 
well-calibrated PL relations in the JHK bands for Milky Way Cepheids, which
agree with the corresponding LMC relations when an improved version of their infrared surface
brightness technique (Gieren, Fouqu{\'e} \& G{\'o}mez, 1997, 1998) is used.

In the {\it Araucaria Project}, started by our group some time ago (Gieren et al. 2005c),
we have conducted surveys for Cepheid variables in a number of galaxies in the Local Group, 
and in the more distant Sculptor Group in order to investigate the effect of environmental properties on
the PL relation, and to improve the accuracy of Cepheids as distance indicators. While we
are discovering Cepheids in optical photometric bands, where these stars are rather easy to detect
due to their relatively large amplitudes and typical light curve shapes (e.g. Pietrzynski et al. 2002, 2004), 
the main goal of the program is
to undertake near-infrared follow-up imaging of selected subsamples of Cepheids
in our target galaxies to obtain accurate reddening information, and thus to obtain more
accurate distances than what is possible from optical (VI) data alone. Near-infrared PL relations
from such Cepheids with existing information on their periods, and V and/or I light curves
can be obtained very economically because it is possible to obtain accurate mean JHK magnitudes for these stars
from just one single-phase observation from the template
light curve approach of Soszynski, Gieren and Pietrzynski (2005). The success
of this approach was recently demonstrated in the case of the Sculptor galaxy NGC 300 (Gieren et al. 2005b).
For this galaxy, a combination of the PL relations obtained in the optical VI and
infrared JK bands has allowed to determine a distance which is practically unaffected
by any remaining uncertainty on reddening. It has also been shown in that paper that
from the combined optical/near-infrared approach
a total uncertainty as small as 3 \% can be obtained for the Cepheid distance  (as measured
relative to the LMC) for such a relatively nearby (2 Mpc) galaxy. 

In the present paper, we are applying the same approach to the Local Group dwarf irregular galaxy IC 1613.
IC 1613 is a very important galaxy in the Araucaria Project because of the very low metallicity of its young stellar
population close to -1.0 dex (Skillman et al. 2003), making it the lowest-metallicity galaxy
in our sample. It is therefore a key object
in our effort to determine the effect of metallicity on the Cepheid PL relation, and on other
stellar distance indicators, like blue supergiant stars (Kudritzki et al. 2003). 
A first survey for Cepheid variables in IC 1613 was carried out by Sandage (1971) who used 
photographic images previously obtained by Baade. Modern work
on the Cepheid PL relation in IC 1613, in the optical V and I bands, has been carried out 
by the OGLE Project (Udalski
et al. 2001) who has discovered many new, previously unknown Cepheids in this irregular galaxy. 
More recently, Antonello et al. (2006) have extended this work to the B and R bands. From the
work of the OGLE group, it could be established that the slope of the PL relation in optical bands is
identical to the slope observed for the more metal-rich LMC Cepheids, arguing for a metallicity-
independent slope of the PL relation. In the near-infrared, a pioneering paper on the
distance of IC 1613 from H-band photometry of 10 Cepheids was published by McAlary et al. (1984)
already 20 years ago;
however, the uncertainty on this distance result was rather large due to the technical difficulties
to obtain accurate IR photometry at those times, and the lack of an accurate calibrating PL relation.
Much more recently, Macri et al. (2001) determined a near-infrared Cepheid distance to IC 1613 from
H-band photometry of four variables obtained with HST/NICMOS. The accuracy of this determination 
suffers, however, from the
very small number of stars used in the PL solution. A main goal of the present study was 
to derive {\it truly accurate near-infrared PL relations for IC 1613}, based on a large number
of well-observed and well-selected Cepheids (see section 3.2.), and this way reduce the current uncertainty
on the distance to IC 1613 to the very small level of 3-5\% we have achieved in our previous study
of NGC 300. 

We have organized this paper in the following way: in section 2,
we describe the observations, reductions and calibration of our data; in section 3, we present
the calibrated infrared mean magnitudes of the Cepheids in our selected fields in IC 1613, and 
determine the distance and reddening; in section 4, we discuss our results; and in section 5, we
summarize the main results of this work, and present some conclusions.

\section{Observations, Data Reduction and Calibration}
\subsection{Optical data}
Our infrared observations of IC 1613 (see next section) were obtained 
about four years after the OGLE-II optical observations (Udalski et al.
2001) of this galaxy. This long gap in time made it necessary to improve the periods
of the Cepheids in order to calculate accurate $<K>$ and $<J>$  
mean magnitudes from single-phase infrared observations with the method of Soszynski et al. (2005).
For this purpose, three new V-band observations 
of IC 1613 with the 1.3 m Warsaw telescope located at Las Campanas 
Observatory were secured in September 2005. This telescope is equipped with a mosaic 8k x 8k detector, 
with a field of view of 35 x 35 arcmin and a scale 
of 0.25 arcsec/pixel. Preliminary data reductions (i.e. debiasing
and flatfielding) were done with the IRAF \footnote{IRAF is distributed by the
National Optical Astronomy Observatories, which are operated by the
Association of Universities for Research in Astronomy, Inc., under cooperative
agreement with the NSF.}  package. The point-spread
function photometry was obtained with the DAOPHOT and ALLSTAR programs, 
in an identical way as described in Pietrzy{\'n}ski et al. (2002). 
Our photometry was then transformed to the standard system using the 
OGLE-II list of carefully calibrated stars in this galaxy (Udalski et
al. 2001).

\subsection{Infrared data}
The near-infrared data presented in this paper were collected with the ESO 
NTT telescope on La Silla, equipped with the
SOFI infrared camera. In the setup we used (Large Field), the field 
of view was 4.9 x 4.9 arcmin, with a scale of 0.288 arcsec / pixel. 
The gain and readout noise were 5.4 e / ADU and 0.4 e, respectively. 

The data were obtained under two observational programs: 
074.D-0318(B), 074.D-0505(B) (PI: Pietrzy{\'n}ski) as part 
of the Araucaria Project. 
Alltogether, six different, slightly overlapping fields were 
observed through J and Ks filters. Their location is shown 
in Figure 1,  and the equatorial coordinates of their centers 
are given in Table 1. 

Single deep J and Ks observations of  our six fields
were obtained under excellent seeing conditions during three different 
photometric nights.  On these nights, we also observed a large number of photometric standard
stars from the UKIRT system (Hawarden et al. 2001). In order to account 
for the rapid sky level variations in the infrared domain, the
observations were performed with a dithering technique.
In the Ks filter, we obtained six consecutive 10 s integrations (DITs) 
at a given sky position and then moved the telescope by about 20 arcsec
to a different random position. Integrations obtained at 65 different 
dithering positions resulted in a total net exposure time of 65 minutes in this
filter, for a given field. In the case of the J filter, in which the sky level variations
are less pronounced than in K, two consecutive 20 s exposures were 
obtained at each of  25 dithering positions, which corresponded to 
a total net exposure of about 17 minutes, for any given field.

Sky subtraction was performed by using a two-step process implying the
masking of stars with the XDIMSUM IRAF package in an analogous manner as described in 
Pietrzy{\'n}ski and Gieren (2002). Then the single images were flatfielded 
and stacked into the final images. PSF photometry was obtained 
using DAOPHOT and ALLSTAR, following
the procedure described in Pietrzy{\'n}ski, Gieren, and Udalski (2002). 
In order to derive the aperture corrections for each frame, about 7-10 
relatively isolated and bright stars were selected, and all
neighbouring stars were removed using an iterative procedure. Finally, 
we measured the aperture magnitudes for the selected stars with the
DAOPHOT program using apertures of 16 pixels. The median from the
differences between the aperture magnitudes obtained this way, and the corresponding PSF
magnitudes, averaged over all selected stars was finally adopted as 
the aperture correction for a given frame. The rms scatter from 
all measurements was always smaller than 0.02 mag.

In order to accurately transform our data to the standard system
a large number (between 8 and 15) of standard stars from the UKIRT 
system (Hawarden et al. 2001) was observed under photometric conditions
at a variety of airmasses, together with our six science fields.
The standard stars were chosen to have colors bracketting the colors of the
Cepheids in IC 1613. The aperture photometry for our standard stars 
was performed with DAOPHOT using the same aperture as for the
calculation of the aperture corrections. Given the relatively 
large number of standard stars we observed on each night, the transformation coefficients 
were derived for each night. The accuracy of the zero points of our photometry was
determined to be about 0.02 mag.

Since our six fields overlap (see Fig. 1), we were able to perform an
internal check of our photometry comparing the magnitudes of stars located 
in the common regions.  In every case, the independently calibrated 
magnitudes agree within 0.02 - 0.03 mag in both, K and J filters.
Unfortunately, we are not aware of any other deep near-infrared JK images
obtained for IC 1613, so an external check of our photometry 
is not possible. However, the magnitudes of the bright stars 
in our fields can be compared with the 2MASS photometry. Fig. 2
presents the difference between our K magnitudes and J-K colors, and the corresponding 
2MASS data for common bright stars. Before calculating these differences, we transformed
our photometry, which 
had been calibrated onto the UKIRT system, to the 2MASS 
system using the equations provided by Carpenter (2001). In spite of the
relatively large uncertainties of the 2MASS data for the fainter stars in Fig. 2, 
it is appreciated that there is
no evident zero point offset either in K or in J-K, supporting the conclusion
that both datasets are well calibrated, within 0.02-0.03 mag.

The pixel positions of the stars were transformed to the equatorial 
coordinate system  using Digital Sky Survey (DSS) images. For this purpose, we used 
the algorithm developed and used in the 
OGLE project (Udalski et al. 1998). The accuracy of our 
astrometric transformations is better than 0.3 arcsec.

\section{Results}
\subsection{The Cepheid mean K and J band magnitudes}
In the six NTT/SOFI fields observed in this project, 39 objects from 
the Cepheid list presented by Udalski et at. (2001) were identified.
It is worth noticing  that most of the (few) long-period OGLE-II Cepheids in IC 1613
are located in our fields. Thanks to the depth of our infrared photometry, we were able 
to detect Cepheids with periods down to about 2 days. 

In Table 2, we present the journal of individual observations of these 39 
Cepheid variables. Most of them were observed only once. However, several 
objects located in the overlapping areas were observed twice.
Before deriving the mean K and J magnitudes of the Cepheids from these data, 
we tried to improve the periods given by Udalski et al. (2001)
using our new three-epoch  optical observations, which are given in Table 3.
For the long-period Cepheids (P larger than about 5 days), 
we could indeed improve the periods with these new data. However, for the variables 
with shorter periods
the time elapsed between two sets of observations 
was too large in order to unambiguously count the number of elapsed
cycles. For these Cepheids, we  adopted the OGLE II periods from Udalski et al. (2001).

The  mean magnitudes were obtained from the template light curve method of Soszy{\'n}ski, Gieren and 
Pietrzy{\'n}ski (2005) which uses the V-band phases of the individual near-IR observations,
and the light curves amplitudes in V and I to calculate the differences of the individual
single-phase magnitudes to the mean magnitudes in J and K.
For a detailed description of this technique, the reader is referred to
this paper. It has been demonstrated by
these authors that the mean K and J magnitudes of Cepheids can be derived from just
one random-phase observation with an accuracy of 0.02-0.03 mag provided that 
high-quality optical and infrared data, and periods are available for the stars.
In Table 4, we present the final intensity mean JK magnitudes of the 39 Cepheids in our fields 
with their estimated uncertainties and their adopted periods. The last
column contains remarks on some of the variables. V2, V6, etc.,
correspond to the numbering system introduced by Sandage (1971).

\subsection{Selection of the final sample}
In Fig. 3, we show the optical V vs. V-I color-magnitude diagram of IC 1613 
obtained from the OGLE II 
data (Udalski et al. 2001) on which the locations of the Cepheids observed 
in the present study are marked. In Fig. 4, we display the
PL relations in the K and J bands which we obtain from the data of all the 39
Cepheids in Table 4.  While the data define tight PL relations in both bands,
there are some objects which clearly deviate from the bulk of the Cepheids in these diagrams,
and which need individual discussion. These stars are indicated with open
circles in Fig. 4.

Star 13682 is most probably not a Cepheid variable (Sandage 1971, 
Antonello et al. 1999). Udalski et al. (2001) supported this conclusion
from the position of this star on the V, V-I CMD  (see Fig. 3), and
its abnormal location on the optical PL relations on which 
13682 appears much brighter than other Cepheids with similar periods.
This proves also true for its near-IR magnitudes in Fig. 4. We therefore exclude this
object for the distance determination.

Besides star 13682, the variables 13709, 11743, 12068,
8173 and 2771 are also very significantly brighter than other Cepheids with
similar periods. The two latter Cepheids with their very short periods 
are almost certainly first overtone pulsators. Due to the detection limit
in our present near-infrared photometry
we would not see fundamental mode pulsators at this very short period 
of about 1.3 days. The three other over-bright Cepheids are 
probably blended by relatively bright stars. These Cepheids are also located above the Cepheid 
PL relations in the V and I bands (Udalski et al. 2001). 
In Fig. 3, variable 11743 appears close to the red edge of the 
instability strip, while the heavily blended Cepheid 13709 lies outside the strip,
supporting the blending hypothesis.
For star 12068, unfortunately no V-I color is available.

The remaining two clearly deviating stars, 10421 and 17473, were already 
classified as Population II Cepheids by Udalski et al. (2001). 
Indeed, these stars are located about 2 magnitudes below the 
IR Cepheid PL relations (see Fig. 4), which fully supports the conclusion
about the Pop. II nature of these objects.

In the light of the arguments presented above, we decided to reject 
all these eight objects from the final sample of Cepheids used for the
distance determination.

Finally, we would like to comment on the Cepheid designated as 7647. 
Udalski et al. (2001) suspected this star to be a heavily blended
Cepheid. Indeed, as can be seen in Fig. 3, this star is located bluewards to 
the instability strip, 
suggesting that this variable is blended with a very bright blue star.
In the infrared, Cepheid 7647 appears with normal flux and colors. 
This finding is consistent with the presence of a blue, unresolved companion star,
which does contaminate the optical, but not the near-infrared photometry.
We therefore retain this Cepheid
in our final list of stars for the distance solution in the infrared. We remark that 
an omission of
this star from the final sample would not significantly alter the results we will 
present below.

The errors of the mean K-band magnitudes for Cepheids with 
log P (days) $<$ 0.5 become large due to a) the relatively low accuracy 
of the K-band photometry for such a faint stars ( $K > 20.5$ mag), and b)
the increasingly uncertain mean magnitude corrections for these stars,
caused by their relatively noisy optical 
light curves, and less accurate periods. We obtained linear regressions to the
PL relations in the J and K bands for the whole sample, including the faintest stars, and 
for the subsamples limited to the Cepheids with log P (days) $>$ 0.5,
finding very good agreement (to better than 1 $\sigma$) between the corresponding solutions. 
However, since the inclusion of the shortest-period Cepheids in the solutions does increase the noise 
significantly, we decided to adopt log P (days) = 0.5  as a lower cut-off period
for our solutions. This way, our final samples in J and K still comprise some
20 Cepheids with excellent photometry, which is sufficient for a very accurate
determination of the distance to IC 1613.

\subsection{Determination of the distance and reddening}
The least-squares fits to the mean magnitudes of the Cepheids from our 
carefully selected final list yield the following 
slopes of the PL relations: $ -3.117 \pm 0.044$ in J and $-3.148 
\pm 0.053 $ in K. The stated errors are $1 \sigma$ uncertainties. 
These values agree very well with the slopes of the PL relations 
for the LMC Cepheids in these bands derived by Persson et al. 
(2004) (-3.153 and -3.261 in J and K, respectively), and are consistent
with the LMC values within the combined uncertainties.
We therefore calulated the zero points of the Cepheid J and K band  PL relations
in IC 1613 by adopting the corresponding slopes from Persson et al.
(2004). This yields the following results: \\

J = -3.153 log P + (22.187 $\pm$ 0.040) \\

K = -3.261 log P + (21.827 $\pm$ 0.045) \\

The adopted linear regressions to our K and J Cepheid data are shown in Fig. 5.
Before calculating, from the determined zero points, the relative distance of IC 1613
with respect to the LMC, we need 
to convert our PL relation zero point magnitudes calibrated onto the UKIRT system
(Hawarden et al. 2001) to the NICMOS system on which the corresponding
LMC zero points were calibrated (Persson et al. 2004). According to 
Hawarden et al. (2001) there are  just zero point offsets between
the UKIRT and NICMOS systems (e.g. no color dependence) in the J and K
filters, which amount to 0.034 and 0.015 mag, respectively. 
After adding these offsets, and assuming an LMC true distance modulus 
of 18.5 mag (see next section for discussion on this assumption), we derived
the following distance moduli for IC 1613: 24.385 (J), 
and  24.306 mag (K) . The corresponding distance moduli in the optical 
V (24.572 mag) and I (24.488 mag) bands had been previously calculated from the 
OGLE-II data by Udalski et al. (2001) adopting the linear LMC Cepheid P-L
relations (Udalski et al. 1999, Udalski 2000).
 
With the values of the distance moduli of IC 1613 derived in four different bands, 
providing the large coverage in wavelength from 0.5-2.2 microns, we can compute 
the reddening, and true distance modulus 
of the galaxy very accurately.
Adopting the extinction law of Schlegel et al. (1998), and following the approach
we have developed in the study of NGC 300 (Gieren et al. 2005b), we fit
a straight line to the relation $(m-M)_{0} =
(m-M)_{\lambda} - A_{\lambda} = (m-M)_{\lambda} - E(B-V) * R_{\lambda} $.
The best least squares fit to this relation yields: \\

$(m-M)_{0} = 24.291 \pm 0.014$   

$ E(B-V) = 0.090 \pm 0.007$

From Fig. 6, it is appreciated that the true distance modulus, and the total reddening
of IC 1613 are indeed very well determined from the available distance moduli in
the different photometric bands.

\section{Discussion}
In the following, we will discuss the various assumptions we made, and 
possible systematic errors which could affect our distance 
determination of IC 1613.

Almost certainly, the largest contribution to our total error budget 
comes from the current uncertainty of the distance to the LMC. Since 
this problem  has been extensively discussed in the recent literature 
(e.g. Benedict et al. 2002; Walker 2003), we will not focus on this discussion here. 
The value of 18.50 mag for the true LMC distance modulus adopted in this
paper assures to have our distance results on the same scale as the results 
from the HST Key Project team (Freedman et al. 2001) 
and our own previous distance studies in the course of the Araucaria Project
(Pietrzynski et al. 2004, Gieren et al. 2004, Gieren et al. 2005b).  

The adopted fiducial slopes of the J- and K-band Cepheid PL relations from
the work of Persson et al. (2004) are based on about 100 LMC Cepheids
with periods bracketting those of the IC 1613 Cepheids used in our present study.
The Persson et al. infrared PL relations clearly represent the most accurate
determination of these relations currently available in the literature. From
the recent work of Gieren et al. (2005a), there is evidence that the infrared
Cepheid PL relations in the Milky Way agree with the corresponding LMC relations,
within the combined $1 \sigma$ uncertainties. In Gieren et al. (2005b), we found
that the slopes of the Persson et al. PL relations in J and K do also provide excellent
fits to the Cepheid near-IR data in NGC 300, with its slightly more
metal-rich young population than the one in the calibrating LMC (Urbaneja et al. 2005).
From the present study of IC 1613, we now see that the slopes of the LMC near-IR PL relations  
give an excellent fit to the metal-poor population of Cepheids in IC 1613, too. This
indicates that on the one hand, use of the Persson et al. LMC PL relations does not
introduce any significant systematic error in our current determination of the IC 1613
distance; on the other hand, it strongly suggests that in the near-infrared domain, as
in the optical domain, the slope of the Cepheid PL relation is independent of metallicity
in the wide range from about -1.0 dex up to solar abundance. This empirical finding is
in good agreement with the model predictions of Bono et al. (1999) who have found that
both, the zero-point and the slope of the K-band PL relation depend only marginally on
metal abundance. They found that the predicted slope in K is 3.19 $\pm$ 0.09 for the LMC,
and 3.27 $\pm$ 0.09 in the SMC, compatible with both the empirical value determined by Persson
et al. (2004) for the LMC, and with a zero change in the slope of the K-band PL relation when going
from LMC (-0.3 dex) metallicity to the SMC (-0.7 dex) metal abundance.

Finally, it is worthwhile to notice that the adoption of the slightly non-linear P-L
relations for Cepheids in the LMC, as advocated by Tammann and Reindl 2002, and more recently Ngeow et
al. 2005, would practically have  no influence on the results presented in
this paper. Indeed, as has already been stated in Ngeow et al. (2005), such an effect
would introduce a change less than 3 percent for the derived
distance modulus, which is in the order of the one $\sigma$ error of our current determination. 
In order to check this out more carefully, we used the Ngeow et al. P-L relations in both optical 
and infrared bands for LMC Cepheids with periods longer
than 10 days as fiducial relations and re-calculated the distance moduli in the VIJK bands. 
This exercise resulted in revised distance moduli to
IC 1613 which in all bands were consistent within one $\sigma$ with our original
results obtained by using the Cepheid P-L relations of Udalski (2000) and 
Persson et al. (2004) for the LMC. The possible non-linearity of the LMC P-L relation, and
the associated slight change of its slope for the long-period Cepheids, is therefore not
a significant problem in the context of our current distance work. Yet, it will be very important
to improve on the slope for the long-period LMC Cepheid P-L relation by using very
accurate and homogeneous new data. We are currently involved in a project to obtain 
such new data in the V and I bands.

The sample of Cepheids used for our present distance determination to IC 1613
is relatively large, making our distance result invulnerable to the problem of an inhomogeneous
filling of the instability strip which is ideally required in such studies.
We suspect that the main reason for the difference of 0.14 mag between 
our current distance result for IC 1613
and the one obtained by Macri et al. (2001) is the small number of Cepheids available
for their study (4), which does not guarantee a homogeneous filling of the instability strip
and can cause a relatively large systematic offset of the derived distance modulus from the
true value. Therefore, we consider our present result as consistent with the HST-based result
of Macri et al. (2001). 
The location of the Cepheids of our final sample in the CMD in Fig. 3 shows that they do indeed cover 
the instability strip quite homogeneously. Moreover, the period range for the PL solution
is very wide and rather uniformly covered with stars-we chose our IR fields in such a way as to optimize
the period distribution of the Cepheids in these fields. 
Applying different cut-off periods 
to our sample (e.g. log P = 0.5, 0.7, 1), we always reproduce the zero point results 
to within one $\sigma$. From this we conclude that our choice for the cut-off period
does not affect our final results in any significant way.

The most important source of uncertainty while using the optical data {\it alone}
is the interstellar reddening. Our present study shows that most of the reddening
to the IC 1613 Cepheids is actually contributed in the galaxy itself, which explains
the overestimation of the distance to IC 1613 in previous studies from optical
data which had only used the very small foreground extinction to IC 1613 to
make the reddening correction.
Using infrared data, and in 
particular K-band photometry where the reddening is by an order of 
magnitude smaller than in the optical bands, the error due to reddening is minimized
to a practically insignificant level of about 0.01 mag. 
From the fact that our new value of E(B-V) yields very consistent 
distances from the PL relations in all optical and infrared bands, we can also 
conclude that the extinction law in IC 1613 is not significantly different 
from the Milky Way law of Schlegel et al. (1998). This is the same conclusion
we had already reached in the case of NGC 300 (Gieren et al. 2005b).

Another contribution to the error budget comes from the effect 
of unresolved companions on the Cepheid magnitudes. The few strongly 
blended Cepheids in our sample were easily detected from their positions on the 
multiband PL relations, and on the CMD, and were discarded from our 
further analysis (see section 3.2.). As we extensively discussed 
in our previous papers (e.g. Gieren et al. 2004, 2005b; Bresolin et al. 2005, 
Pietrzynski et al. 2004) the blending effect was found to be very small in the cases of 
NGC 300, and of NGC 6822. In the paper of Bresolin et al. (2005), we were able
to demonstrate from HST/ACS images that those Cepheids in NGC 300 which we had
identified as strongly blended in the ground-based photometry were indeed the ones
with the brightest nearby companions. In that paper, it was shown that the
effect of unresolved companion stars on the Cepheids which constituted the
final sample was less than 2 percent. Given that IC 1613 is located at less than
half the distance of NGC 300, and has a much smaller stellar density, it is
reasonable to assume that the effect of blending due to unresolved companion
stars on its distance is even smaller than in the case of NGC 300, and
does not contribute in a significant way to the systematic uncertainty of
our present result. 

While it now seems well established that the slopes of the 
Cepheid P-L relations in optical and near-infared bands
do not depend, within our current detection sensitivity, on metallicity over a very broad range of 
this parameter ( -1 $<$ [Fe/H] $<$ 0 ; see previous discussion),
a possible metallicity dependence 
of their {\it zero points} is still under discussion (Sakai et al. 
2004; Storm et al. 2004; Pietrzynski et al. 2004; Pietrzynski and Gieren 2005).
In particular, due to the fact that up to now very few galaxies have been {\it exhaustively}
 surveyed for Cepheids in the infrared,
it is currently not possible to draw any firm conclusion about the 
potential variation of the infrared PL relation zero points with 
metallicity. Soon, once the data for all target galaxies 
observed in the course of the Araucaria Project will be analyzed, 
we should be able to put tighter constraints on this open question,
and if needed calibrate the metallicity dependence of PL relation zero 
points in both optical and infrared domains with high precision. 

\section{Summary and conclusions}

We have measured accurate NIR magnitudes in the J and K bands for 39 Cepheid variables
in the Local Group galaxy IC 1613 with well-determined periods and optical (VI)
light curves. Mean magnitudes in J and K were derived for these variables using
the single-phase approach of Soszy{\'n}ski et al. (2005). After carefully cleaning 
the Cepheid list from blended objects, Population II variables and overtone pulsators,
we have determined accurate PL relations. Fits to these observed relations were made
using the slopes of the LMC relations determined by Persson et al. (2004), which gave an excellent
representation of the IC 1613 data, providing for the first time solid evidence
that the slope of the Cepheid PL relation is independent of metallicity down to
the low metallicity of -1.0 dex of the IC 1613 young population in the near-infrared
domain, too. This is in agreement with the theoretical predictions of Bono et al. (1999).
 By combining the zero points of the J and K band PL relations in our
study with the ones derived by Persson et al. for the LMC, we derive relative distance
moduli of IC 1613 with respect to the LMC in both bands. Combining these infrared
moduli with the distance moduli previously derived by Udalski et al. (2001) in V
and I, we determine the total (average) reddening of the Cepheids in IC 1613, and
the true distance modulus of this galaxy with an unprecedented accuracy. For the
reddening, we find E(B-V) = 0.090 $\pm$ 0.007 mag, and for the true distance modulus
of IC 1613 from our multiwavelength approach we obtain 24.291 $\pm$ 0.014 mag (random error).
As in the case of our study of NGC 300 with the same method, we find evidence that there
is a significant contribution to the total reddening from dust absorption {\it intrinsic}
to IC 1613, which had been neglected in the previous Cepheid distance work on
this galaxy. The excellent fit of the distance moduli to the assumed galactic extinction law
suggests that the interstellar extinction in this small irregular galaxy follows closely
the galactic law.

We show that our derived Cepheid distance is very insensitive to systematic uncertainties
caused by the fiducial PL relations used in our fits, possible inhomogeneous filling
of the instability strip by our Cepheid sample, and problems with blending of the variables.
Any remaining influence of the uncertainty of reddening on our distance result is
negligible. All these possible sources of error contribute less to the total systematic
uncertainty of our result than the two dominant sources of error, which are the zero
points of our JK photometry ($\pm$ 0.03 mag), and the distance to the LMC, which we have
{\it adopted} as 18.50 mag, and whose current uncertainty seems in the order of $\pm$ 0.10 mag. 

Our distance determination for IC 1613 is in reasonable agreement with the previous determination
of Macri et al. (2001) from HST H-band photometry of four Cepheids, 24.43 $\pm$ 0.08 mag. 
We attribute the 0.14 mag difference mainly to the small number of stars available to Macri et al.
in their study.
Our new distance determination is also in very good agreement with the very early infrared
work of McAlary et al. (1984); these authors had obtained a distance modulus of 24.31 $\pm$ 0.12
from H-band data of 10 Cepheids in IC 1613. A change of their assumed reddening of 0.03 mag
to our larger value found in this study still produces excellent agreement of their result
with ours. 

The distance to IC 1613 was also determined  by Dolphin et al. (2001) using HST data, and employing
a number of distance  indicators (TRGB, red clump stars, RR Lyrae stars, Cepheids). We note
that their Cepheid sample was very small, which can clearly lead to spurious results.
Udalski et al. 2001 have observed an order of magnitude larger sample of
Cepheids in this galaxy and  showed that all these different distance indicators yield
consistent distances to this galaxy. Those distance measurements are all in very good
agreement with the distance of IC 1613 obtained in this study if the revised reddening of 0.09 mag
found in this study, and a LMC true distance modulus of 18.5 mag are assumed. 

As a final conclusion, we have produced in this work a determination of the Cepheid distance
to IC 1613 whose random error is of the order of 1\%, and estimated systematic error (excluding
the uncertainty of the adopted LMC distance) is in the order of 2\%. This accuracy, when
combined with distance determinations of similar accuracy we pretend to obtain for most of the other
galaxies of the Araucaria Project, should enable us to pin down the metallicity dependence
of the PL relation zero points in the different optical and near-infrared photometric bands with
the 1-2\% accuracy needed to produce a significant improvement in the determination of
the Hubble constant from distance determinations to galaxies in the nearby field from
their Cepheid populations, which is the approach used in the HST Key Project of Freedman et al. (2001). 
Our work in the Araucaria Project should therefore strongly contribute, in the very near future,
to make best use of the past work of the Key Project team.

\acknowledgments
We would like to thank the anonymous referee for his interesting suggestions 
which helped to improved this paper. 
We gratefully acknowledge the generous allocation of observing time
by ESO to our distance scale projects. We also appreciate the excellent staff
support at the telescope at ESO/La Silla where these data were obtained.
WG and GP gratefully acknowledge financial support for this
work from the Chilean Center for Astrophysics FONDAP 15010003. 
Support from the Polish KBN grant No 2P03D02123 and BST grant for 
Warsaw University Observatory is also acknowledged.

\begin{figure}[p] 
\vspace*{18cm}
\includegraphics{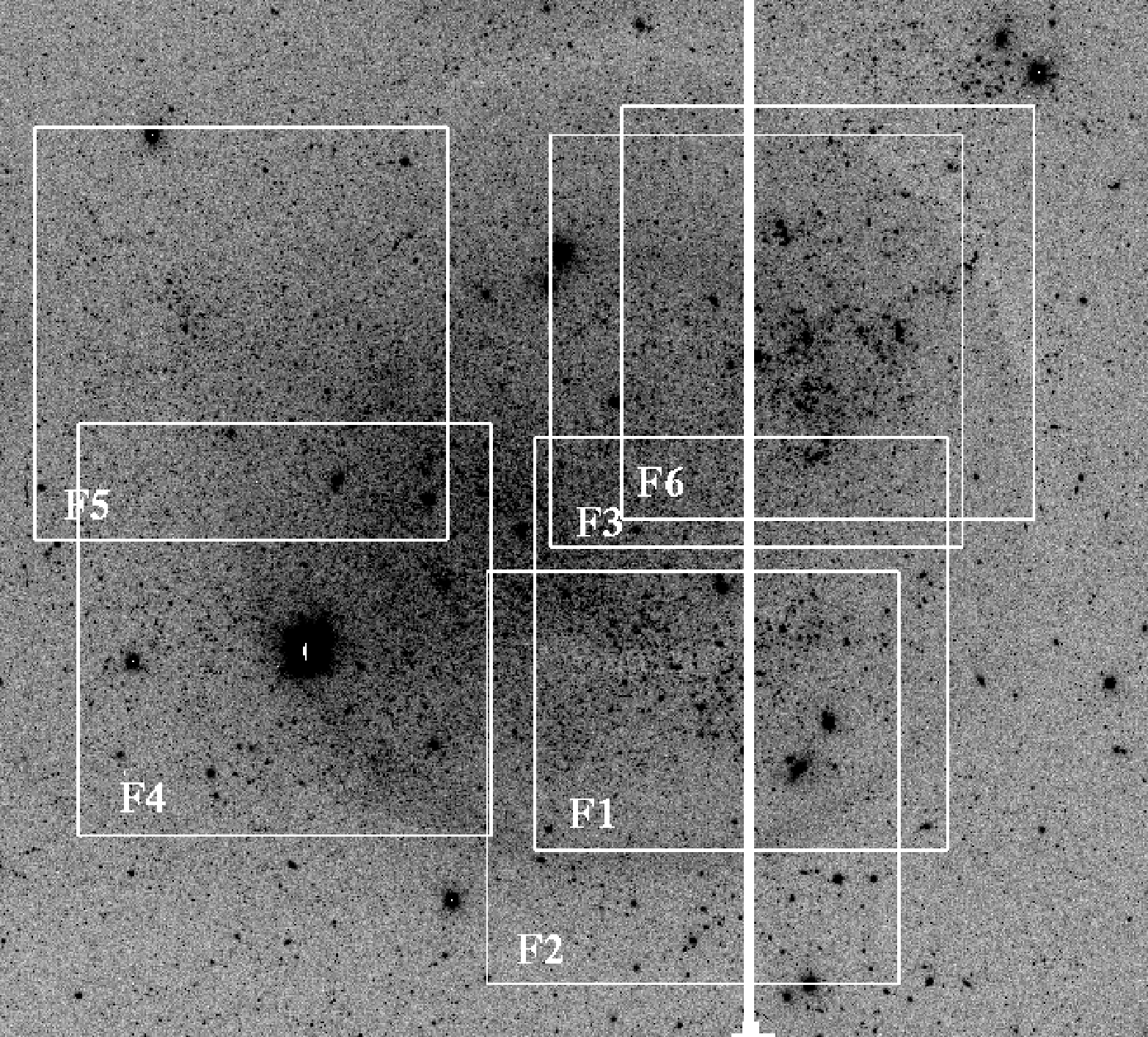} 
\caption{The location of the observed NTT fields in IC1613 on a V-band image of
this galaxy, taken with the Warsaw 1.3-m telescope on Las Campanas.}
\end{figure}

\begin{figure}[p]
\vspace*{18cm}
\includegraphics{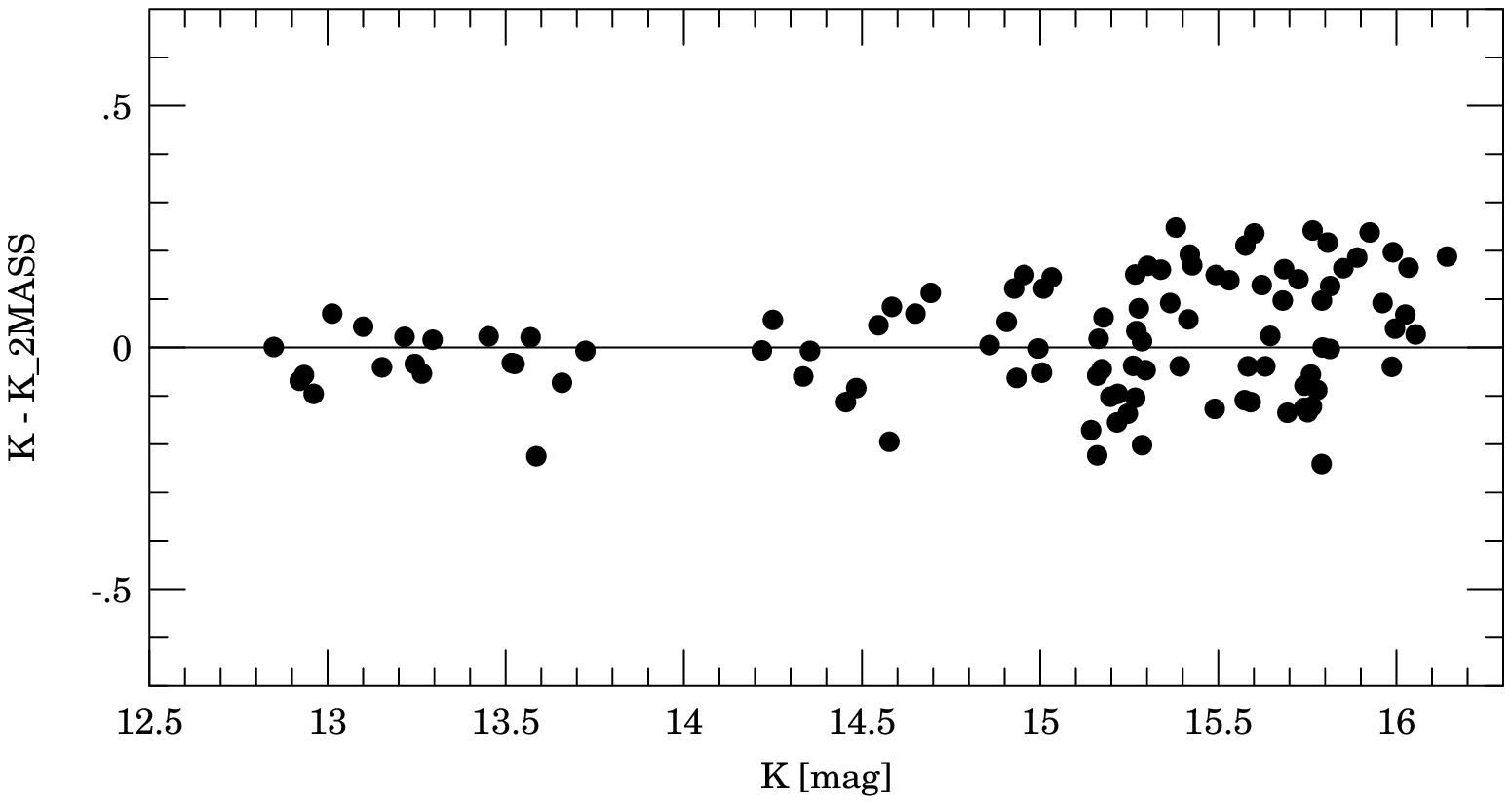}
\includegraphics{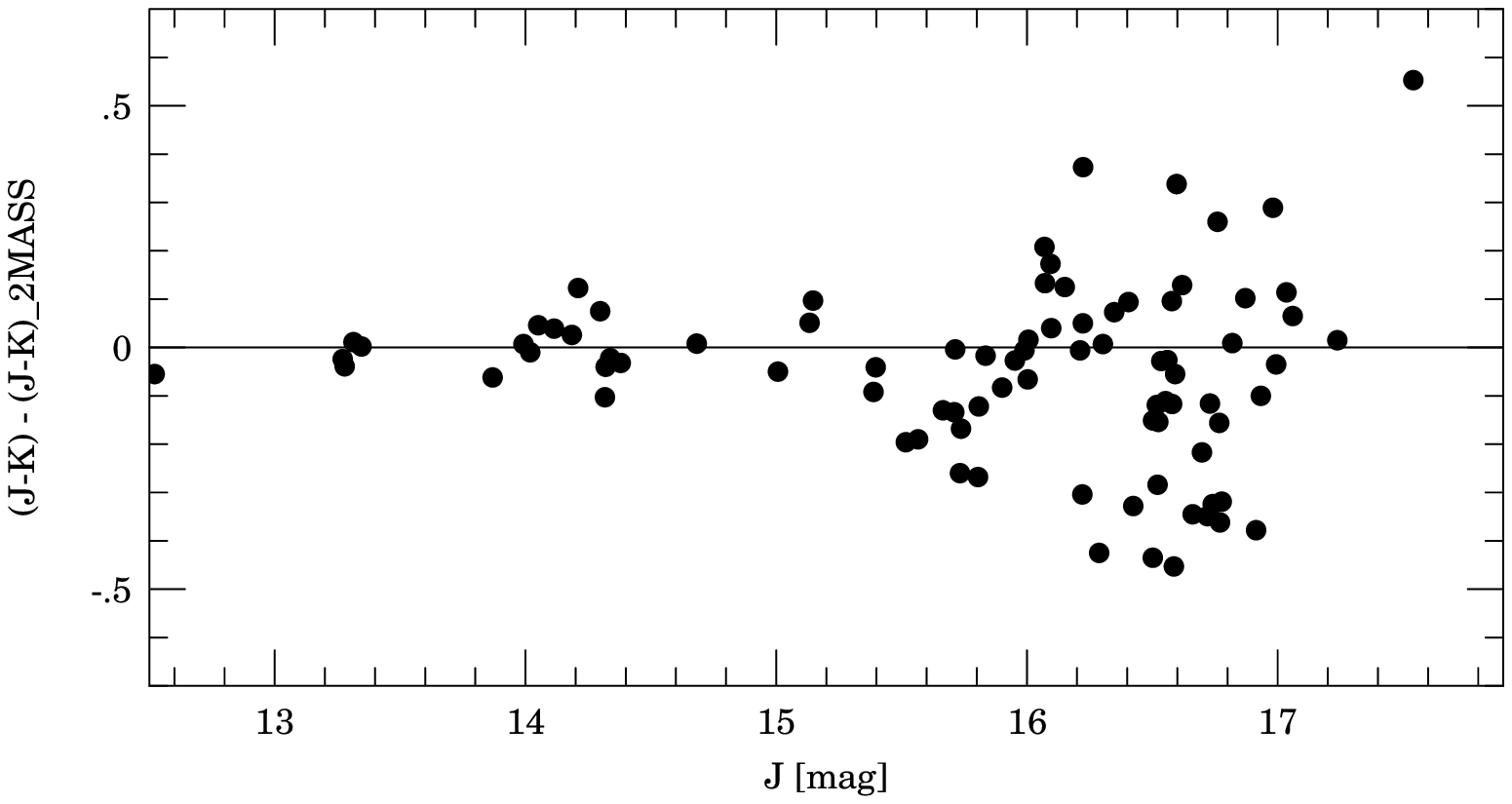}
\vspace{-3cm}
\caption{Comparison of our present photometry with the 2MASS data. In spite of the 
relatively large scatter towards the fainter magnitudes caused by the low accuracy 
of the 2MASS photometry
of faint stars, no evident zero point offset either in K or in 
J-K is present.}
\end{figure}

\begin{figure}[p]
\includegraphics{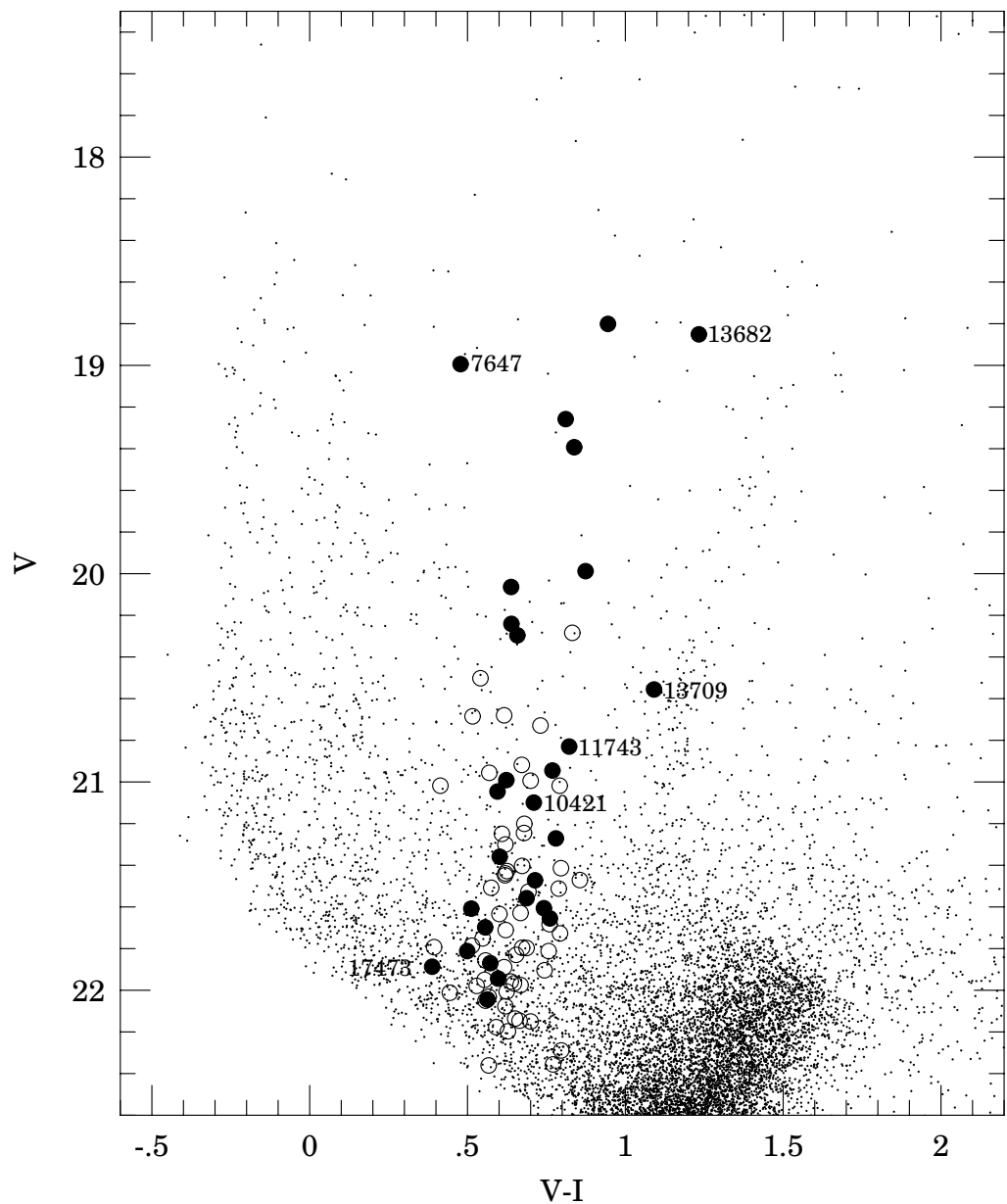}   
\vspace{18cm}
\caption{The V, V-I color-magnitude diagram for the stars in IC 1613 based on 
the OGLE-II data (Udalski et al. 2001). It is demonstrated that the 
Cepheids observed in infrared bands and used for distance determination in the present study  
(stars marked with filled circles, without numbers) uniformly cover the instability 
strip. Cepheids with filled circles and given OGLE-II identifications are discussed in the text. 
OGLE-II Cepheids not covered in our infrared study are marked with open circles.  }
\end{figure}

\begin{figure}[p]
\includegraphics{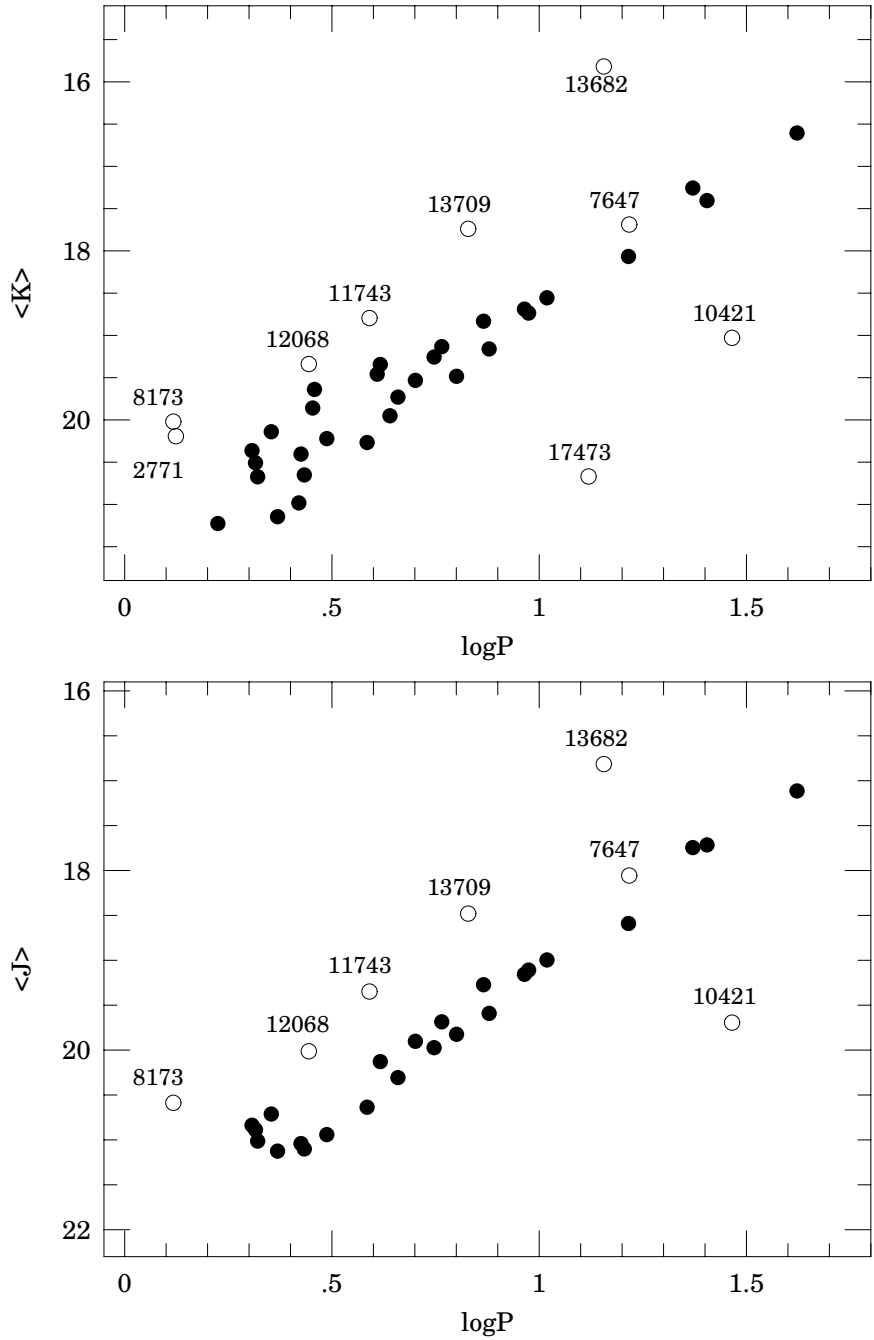}
\vspace{18cm} 
\caption{Period-luminosity relations in the K and J bands, for all Cepheids presented in Table 
3. The stars labelled with their OGLE-II identifications are mostly outliers and are
discussed in the text.}                                                  
\end{figure}

\begin{figure}[p]
\includegraphics{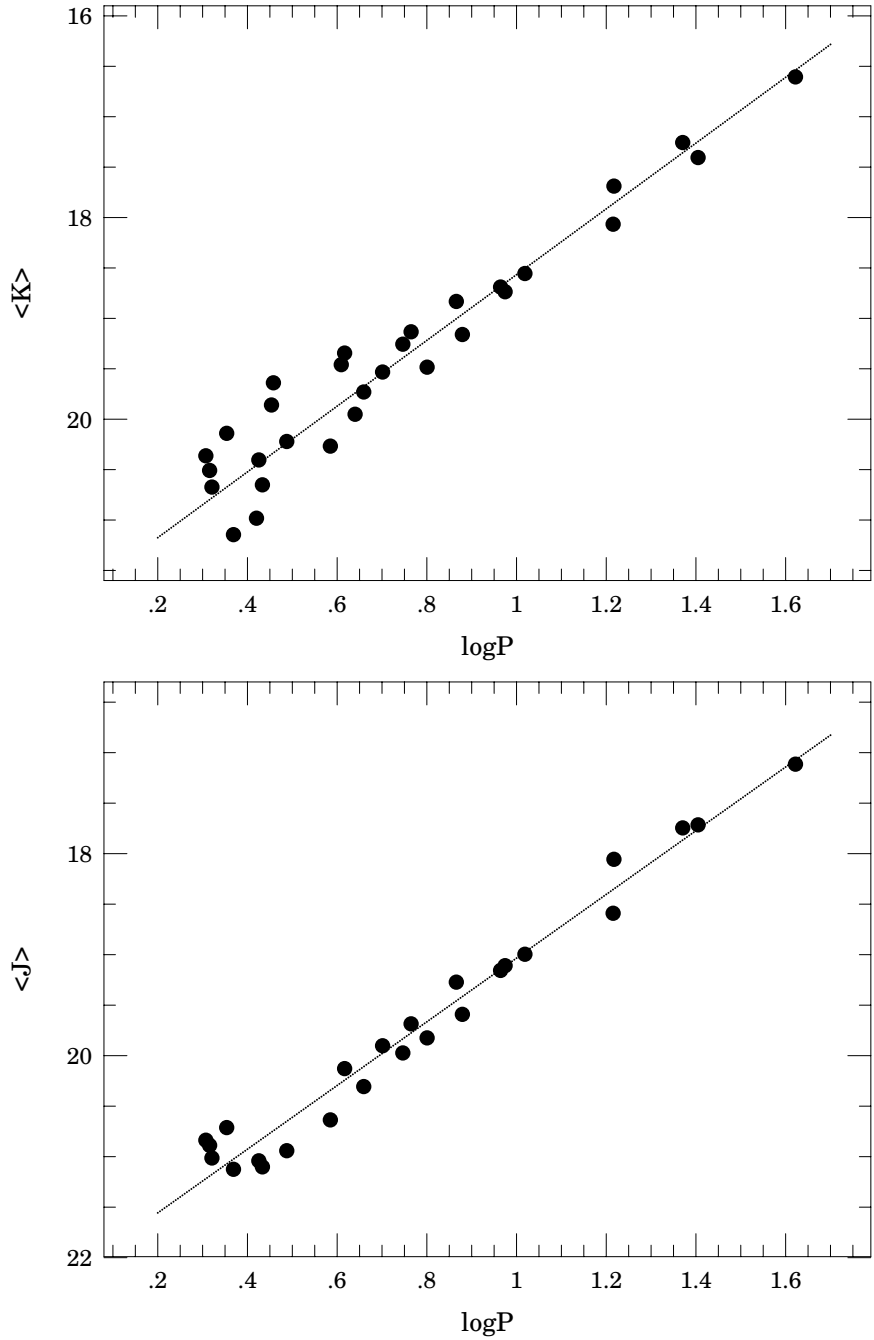}
\vspace{18cm} 
\caption{The final, adopted period-luminosity relations for IC 1613 Cepheids in the J and K bands. The 
slope of the relations were adopted from the LMC Cepheids (Persson et al. 2004). They give excellent
fits to the observed PL relations in IC 1613.
}                                                  
\end{figure}

\begin{figure}[p]
\includegraphics{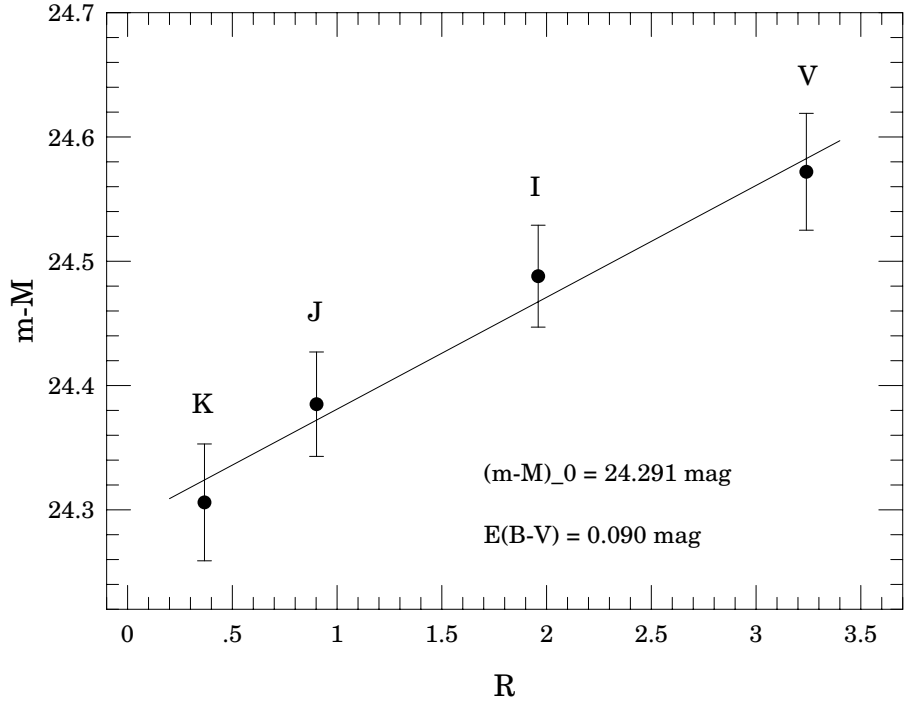}
\vspace{10cm}  
\caption{Apparent distance moduli to IC 1613 as derived in different photometric bands,
plotted against the ratio of total to selective extinction as adopted from
the Schlegel et al. reddening law. The intersection and 
slope of the best-fitting line give the true distance modulus and 
reddening, respectively. The data in this diagram suggest that the galactic reddening law
is a very good approximation for IC 1613 as well.}
\end{figure}

\clearpage

\begin{deluxetable}{c c c}
\tablewidth{0pc}
\tablecaption{Coordinates of the Centers of the Observed SOFI/NTT Fields in IC 1613}
\tablehead{ Field & \colhead{RA} & \colhead{DEC} }
\startdata
  C1   &  01:04:58.9  &  02:05:48.5  \nl
  C2   &  01:04:56.6  &  02:04:13.2   \nl
  C3   &  01:04:59.6  &  02:09:23.9   \nl
  C4   &  01:04:37.2  &  02:05:58.8   \nl      
  C5   &  01:04:35.1  &  02:09:29.4   \nl      
  C6   &  01:05:03.0  &  02:09:44.4   \nl
\enddata
\end{deluxetable}

\clearpage

\begin{deluxetable}{ccccccc}
%\rotate
\tablewidth{0pc}
\tablecaption{Journal of the Individual J and K Observations of IC 1613
Cepheids}
\tablehead{ \colhead{ID} & HJD (J) &  J & $\sigma_{\rm J}$ & HJD (K) & K &  $\sigma_{\rm K}$}
\startdata
11446 &  53215.86015 & 16.997 & 0.009 &  53215.80516 & 16.492 & 0.009 \\
11446 &  53315.64411 & 17.400 & 0.016 &  53315.58668 & 16.822 & 0.016 \\
10421 &  53315.64411 & 19.551 & 0.060 &  53315.58668 & 18.877 & 0.071 \\
 1987 &  53370.57443 & 17.938 & 0.019 &  53370.52660 & 17.593 & 0.024 \\
  736 &  53315.72831 & 17.833 & 0.018 &  53315.67116 & 17.260 & 0.018 \\
 7647 &  53370.57443 & 18.144 & 0.020 &  53370.52660 & 17.716 & 0.027 \\
13738 &  99999.99999 & 99.999 & 9.999 &  53215.88146 & 18.165 & 0.032 \\
13738 &  53315.56228 & 18.763 & 0.043 &  53315.50158 & 18.249 & 0.046 \\
13682 &  99999.99999 & 99.999 & 9.999 &  53215.88146 & 15.763 & 0.007 \\
13682 &  53315.56228 & 16.760 & 0.010 &  53315.50158 & 15.751 & 0.007 \\
\enddata
\end{deluxetable}

\clearpage
\begin{deluxetable}{cccc}
%\rotate
\tablewidth{0pc}
\tablecaption{Journal of the Individual V band Observations of IC 1613
Cepheids}
\tablehead{ \colhead{ID} & HJD & V  & $\sigma_{\rm V}$}
\startdata
11446 & 53620.76797 & 18.385 &  0.009 \nl
11446 & 53621.75534 & 18.422 & 0.008 \nl
11446 & 53621.82741 & 18.412 & 0.008 \nl
10421 & 53620.76797 & 21.632 & 0.079 \nl
10421 & 53621.75534 & 21.648 & 0.069 \nl
 1987 & 53620.76797 & 19.764 & 0.019 \nl
 1987 & 53621.75534 & 19.815 & 0.018 \nl
 1987 & 53621.82741 & 19.829 & 0.019 \nl
  736 & 53620.76797 & 19.593 & 0.017 \nl
  736 & 53621.75534 & 19.671 & 0.018 \nl
  736 & 53621.82741 & 19.675 & 0.023 \nl
 7647 & 53620.76797 & 19.110 & 0.014 \nl
 7647 & 53621.75534 & 19.080 & 0.012 \nl
 7647 & 53621.82741 & 19.100 & 0.012 \nl
13738 & 53620.76797 & 20.035 & 0.020 \nl
13738 & 53621.75534 & 20.115 & 0.021 \nl
13738 & 53621.82741 & 20.109 & 0.023 \nl
\enddata
\end{deluxetable}

\clearpage

\begin{deluxetable}{cccccccccccc}
\tablewidth{0pc}
\tablecaption{Final Intensity Mean J and K Magnitudes of IC 1613 Cepheids}
\tablehead{ \colhead{OGLE ID} & P & $\log{P}$ & $\langle{J}\rangle$ & $\sigma_J$ & $\langle{K}\rangle$ &
$\sigma_K$& remarks}
\startdata
11446 &  41.87 &   1.62194 &  17.114 &   0.009 &  16.605 &   0.009& V20\\
10421 &  29.19 &   1.46529 &  19.694 &   0.060 &  19.029 &   0.071&PII, V47\\
1987 &  25.398 &   1.40480 &  17.715 &   0.019 &  17.405 &   0.024&V11\\
736 &  23.469 &   1.37049 &  17.745 &   0.018 &  17.256 &   0.018&V2\\
7647 &  16.488 &   1.21716 &  18.056 &   0.020 &  17.688 &   0.027&blend\\
13738 &  16.420 &   1.21537 &  18.590 &   0.043 &  18.066 &   0.028&V18\\
13682 &  14.317 &   1.15585 &  16.815 &   0.010 &  15.818 &   0.005&not Cepheid, V39\\
17473 &  13.154 &    1.11906 &  99.999 &   9.999 &  20.669 &   0.251 &PII\\
7664 &  10.4390 &   1.01866 &  18.996 &   0.030 &  18.555 &   0.048&V16\\
926 &   9.4286 &   0.97445 &  19.109 &   0.028 &  18.736 &   0.036&V6\\
879 &   9.2130 &   0.96440 &  19.156 &   0.046 &  18.689 &   0.054&V25\\
13808 &   7.572 &   0.87921 &  19.591 &   0.092 &  19.160 &   0.058&\\
13759 &   7.3403 &   0.86571 &  19.272 &   0.074 &  18.832 &   0.112&V7\\
13709 &   6.741 &   0.82872 &  18.480 &   0.041 &  17.739 &   0.020&blend\\
5037 &   6.3175 &   0.80055 &  19.824 &   0.065 &  19.484 &   0.127&\\
11604 &   5.8191 &   0.76486 &  19.685 &   0.051 &  19.133 &   0.097&\\
13780 &   5.5771 &   0.74641 &  19.973 &   0.137 &  19.256 &   0.069&V9\\
11831 &   5.0269 &   0.70130 &  19.902 &   0.063 &  19.532 &   0.111&\\
8146 &   4.5630 &   0.65925 &  20.306 &   0.075 &  19.730 &   0.122&\\
14287 &   4.365 &    0.63998 &  99.999 &   9.999 &  19.951 &   0.142 &\\
12109 &   4.1364 &   0.61662 &  20.128 &   0.079 &  19.344 &   0.093&\\
13784 &   4.0657 &    0.60914 &  99.999 &   9.999 &  19.459 &   0.096 &V10\\
11743 &   3.8953 &   0.59054 &  19.348 &   0.035 &  18.795 &  0.047&blend, V53\\
8127 &   3.8444 &   0.58483 &  20.636 &   0.097 &  20.267 &   0.159&\\
2240 &   3.0733 &   0.48760 &  20.941 &   0.138 &  20.221 &   0.147&V35\\
18349 &   2.8700 &    0.45788 &  99.999 &   9.999 &  19.639 &   0.102 &V29\\
19024 &   2.8418 &    0.45359 &  99.999 &   9.999 &  19.859 &   0.154 &\\
12068 &   2.781 &   0.44420 &  20.013 &   0.060 &  19.339 &   0.091&blend\\
2760 &   2.7123 &   0.43334 &  21.101 &   0.127 &  20.651 &   0.203&\\
10804 &   2.6629 &   0.42535 &  21.041 &   0.137 &  20.404 &   0.197&V48\\
12526 &   2.6310 &    0.42012 &  99.999 &   9.999 &  20.982 &   0.358 &\\
7322 &   2.3378 &   0.36881 &  21.125 &   0.141 &  21.145 &   0.328&\\
6128 &   2.2578 &   0.35369 &  20.712 &   0.114 &  20.140 &   0.174&\\
8782 &   2.0930 &   0.32077 &  21.013 &   0.137 &  20.673 &   0.255&\\
5996 &   2.0682 &   0.31559 &  20.888 &   0.183 &  20.508 &   0.284& V60\\
2389 &   2.0286 &   0.30720 &  20.837 &   0.115 &  20.363 &   0.175&\\
13481 &   1.678 &    0.22479 &  99.999 &   9.999 &  21.226 &   0.392 &\\
2771 &   1.3290 &    0.12352 &  99.999 &   9.999 &  20.193 &   0.150 &FO\\
8173 &   1.3103 &   0.11737 &  20.587 &   0.099 &  20.020 &   0.130&FO\\
\enddata
\end{deluxetable}

\clearpage

\begin{deluxetable}{cccccc}
\tablewidth{0pc}
\tablecaption{Reddened and Extinction-Corrected Distance Moduli for IC 1613 in Optical and Near-Infrared Bands}
\tablehead{ \colhead{Band} & $V$ & $I$ & $J$ & $K$ & $E(B-V)$ }
\startdata
 $m-M$                &   24.572 &  24.488 &  24.385 &  24.306 &   --   \nl
 ${\rm R}_{\lambda}$  &   3.24   &  1.96   &  0.902  &  0.367  &   --   \nl
$(m-M)_{0}$           &   24.277 &  24.309 &  24.302 &  24.273 &  0.090 \nl
\enddata
\end{deluxetable}

\end{document}